
\catcode`@=11

\newskip\ttglue

\font\twelverm=cmr12 \font\twelvebf=cmbx12
\font\twelveit=cmti12 \font\twelvesl=cmsl12

\font\ninerm=cmr9
\font\eightrm=cmr8
\font\sixrm=cmr6
\font\eighti=cmmi8   \skewchar\eighti='177
\font\sixi=cmmi6     \skewchar\sixi='177
\font\ninesy=cmsy9   \skewchar\ninesy='60
\font\eightsy=cmsy8  \skewchar\eightsy='60
\font\sixsy=cmsy6    \skewchar\sixsy='60
\font\eightbf=cmbx8
\font\sixbf=cmbx6
\font\eighttt=cmtt8  \hyphenchar\eighttt=-1
\font\eightit=cmti8
\font\eightsl=cmsl8

\def\smalltype{\def\rm{\fam0\eightrm}
 			\textfont0=\eightrm  \scriptfont0=\sixrm  \scriptscriptfont0=\fiverm
 			\textfont1=\eighti   \scriptfont1=\sixi   \scriptscriptfont1=\fivei
 			\textfont2=\eightsy  \scriptfont2=\sixsy  \scriptscriptfont2=\fivesy
 			\textfont3=\tenex    \scriptfont3=\tenex  \scriptscriptfont3=\tenex
    \textfont\itfam=\eightit  \def\it{\fam\itfam\eightit}
	   \textfont\slfam=\eightsl  \def\sl{\fam\slfam\eightsl}
	   \textfont\ttfam=\eighttt  \def\tt{\fam\ttfam\eighttt}
    \textfont\bffam=\eightbf  \scriptfont\bffam=\sixbf
        \scriptscriptfont\bffam=\fivebf  \def\bf{\fam\bffam\eightbf}
    \tt  \ttglue=.5em plus.25em minus.15em
    \normalbaselineskip=9pt
    \setbox\strutbox=\hbox{\vrule height7pt depth2pt width0pt}
    \let\sc=\sixrm  \let\big=\eightbig  \normalbaselines\rm}
\def\eightbig#1{{\hbox{$\textfont0=\ninerm\textfont2=\ninesy
    \left#1\vbox to6.5pt{}\right.\n@space$}}}

\def\medtype{\def\rm{\fam0\tenrm}
 			\textfont0=\tenrm  \scriptfont0=\sevenrm  \scriptscriptfont0=\fiverm
 			\textfont1=\teni   \scriptfont1=\seveni   \scriptscriptfont1=\fivei
 			\textfont2=\tensy  \scriptfont2=\sevensy  \scriptscriptfont2=\fivesy
 			\textfont3=\tenex    \scriptfont3=\tenex  \scriptscriptfont3=\tenex
    \textfont\itfam=\tenit  \def\it{\fam\itfam\tenit}
	   \textfont\slfam=\tensl  \def\sl{\fam\slfam\tensl}
	   \textfont\ttfam=\tentt  \def\tt{\fam\ttfam\tentt}
    \textfont\bffam=\tenbf  \scriptfont\bffam=\sevenbf
        \scriptscriptfont\bffam=\fivebf  \def\bf{\fam\bffam\tenbf}
    \tt  \ttglue=.5em plus.25em minus.15em
    \normalbaselineskip=12pt
    \setbox\strutbox=\hbox{\vrule height8.5pt depth3.5pt width0pt}
    \let\sc=\eightrm  \let\big=\tenbig  \normalbaselines\rm}

\def\bigtype{\let\rm=\twelverm \let\bf=\twelvebf
\let\it=\twelveit \let\sl=\twelvesl \rm}

\def\footnote#1{\edef\@sf{\spacefactor\the\spacefactor}#1\@sf
    \insert\footins\bgroup\smalltype
    \interlinepenalty100 \let\par=\endgraf
    \leftskip=0pt  \rightskip=0pt
    \splittopskip=10pt plus 1pt minus 1pt \floatingpenalty=20000
  \vskip4pt\noindent\hskip20pt\llap{#1\enspace}
\bgroup\strut\aftergroup\@foot\let\next}
\skip\footins=12pt plus 2pt minus 4pt \dimen\footins=30pc

\def\bigfont{\magnification=1200 \baselineskip=20pt}

\def\e{\epsilon}

\def\cl#1{\centerline{#1}}
\def\clbf#1{\centerline{\bf #1}}

\def\is#1{{\narrower\smallskip\noindent#1\smallskip}}

\long\def\myname{\medskip
\cl{Kiho Yoon}
\cl{Department of Economics, Korea University}
\cl{145 Anam-ro, Seongbuk-gu, Seoul, Korea 02841}
\cl{ \tt kiho@korea.ac.kr}
\cl{\tt https://kihoyoon.github.io}
\medskip}

\def\ve{\vfill\eject}

\def\frac#1#2{{#1 \over #2}}
\def\Re{I\!\!R}

\newcount\sectnumber
\def\Section#1{\global\advance\sectnumber by 1 \bigskip
           \noindent{\bigtype {\bf \the\sectnumber  \ \ \ #1}} \medskip}

\def\prop#1{\medskip\noindent {\bf Proposition #1.} \it}
\def\lemma#1{\medskip\noindent {\bf Lemma #1.} \it}

\def\eg#1{\medskip\noindent {\bf Example #1.}}

\def\ok{\smallskip \rm}

\def\pf{\medskip\noindent Proof: \/}
\def\pfo#1{\medskip\noindent {\bf Proof of #1:\/}}
\def\endpf{\hfill {\it Q.E.D.} \smallskip}

\def\appx{\bigskip {\bigtype \clbf{Appendix}} \medskip}

\newcount\notenumber
\def\note#1{\global\advance\notenumber by 1
            \footnote{$^{\the\notenumber}$}{#1} \tenrm}

\def\ref{\bigskip \centerline{\bf REFERENCES} \medskip}

\def\jet{{\it Journal of Economic Theory\/ }}
\def\aer{{\it American Economic Review\/ }}

\def\res{{\it Review of Economic Studies\/ }}

\def\ej{{\it Economic Journal\/ }}
\def\et{{\it Economic Theory\/ }}
\def\qje{{\it Quarterly Journal of Economics\/ }}
\def\rje{{\it Rand Journal of Economics\/ }}

\def\jle{{\it Journal of Law and Economics\/ }}

\def\wp{{\it working paper\/ }}

\def\paper#1#2#3#4#5{\noindent\hangindent=20pt#1 (#2), ``#3,'' #4, #5.\par}
\def\wp#1#2#3#4{\noindent\hangindent=20pt#1 (#2), ``#3,'' #4.\par}
\def\book#1#2#3#4{\noindent\hangindent=20pt#1 (#2), {\it #3,} #4.\par}

\bigfont

{ \ }

\vskip 1cm

{\bigtype
\clbf{Uniform price auction with quantity constraints}
}

\vskip 1cm
\bigskip

\myname

\vskip 0.5cm

\clbf{Abstract}
\is{\baselineskip=12pt We study the equilibria of uniform price auctions where many asymmetric bidders have flat demands up to their respective quantity constraints. We present an iterative procedure that systematically finds an equilibrium outcome as well as an ascending auction that has this outcome as a dominant strategy equilibrium outcome. Demand reduction and low price equilibrium may occur since it is advantageous for a bidder to give up some of his/her demand and get the remaining demand at a low price rather than to get his/her entire demand at a higher price. We show that a low price equilibrium is the only possible equilibrium when no bidder's quantity constraint is large enough to cover the supply.}
\smallskip

\is{\baselineskip=12pt Keywords: capacity constraint, ascending auction, procurement, demand reduction, low price equilibrium}
\smallskip

\is{\baselineskip=12pt  JEL Classification: C72; D44; D82}

\ve

\Section{Introduction}

We study the equilibria of uniform price auctions where many asymmetric bidders have flat demands up to their respective quantity constraints. Though not analyzed extensively in the literature, this class of auction environments is common in real-world markets such as wholesale electricity markets, spectrum auctions, and securities markets. The predominant auction format in the European electricity market is the uniform price auction, also known as the `pay-as-clear' auction (ACER, 2022, p. 19). The same is broadly true for electricity markets in the United States. `From the spectrum auctions of 1994 onward, virtually all of these auctions have been uniform price auctions...' (Milgrom, 2004, p. 255). The US treasury securities, as well as most private securities, are largely sold via the uniform price auction, also known as the `single-price' auction (Malvey and Archibald, 1998; Horta\c{c}su {\it et al.\/}, 2018, p. 148).

Quantity constraints may arise due to demand/supply conditions or regulations. For example, buyers in an auction might be retailers with limited final consumer demand, possibly due to low regional population density. Similarly, sellers in a procurement auction (such as electricity companies in day-ahead markets) may have constant marginal costs up to their capacity limits.  In spectrum auctions, the government often imposes spectrum caps that limit the quantity of spectrum a single bidder may obtain so as to prevent excessive market concentration and enhance competitiveness.\note{See Cramton {\it et al.\/} (2011) for a detailed discussion of spectrum auctions.} In treasury auctions and equity initial public offerings (IPOs), large investors may not acquire more than a certain percentage of the total shares.

We first present an iterative procedure for finding an equilibrium outcome of the uniform price auction when there are many quantity-constrained bidders. This outcome is a Nash equilibrium outcome when bidders' valuations for the good and the quantity constraints are common knowledge among bidders.\note{This assumption may not be unreasonable for markets in which the same set of well-established bidders interact frequently.} Moreover, this is the dominant strategy equilibrium outcome of an ascending auction when bidders have private information about their respective valuations. Our ascending auction is a variant of the English auction that incorporates bidders' quantity constraints. In this auction, the clock showing the current price increases continuously over time and bidders decide when to drop out. In addition, the auction keeps track of the provisional price, which is set to zero or to the reserve price initially. When a bidder drops out, the clock stops temporarily and the auctioneer checks whether the aggregate demand of remaining bidders is greater than or equal to the supply. If the former is less than the latter, then the auction ends and the final price is the provisional price. The remaining bidders get their respective demands and the bidder who just dropped out gets the residual supply. Otherwise, the provisional price is updated to the current price just stopped: (i) if the aggregate demand is equal to the supply, the auction ends and the final price is set to the provisional price, which has just been updated. The remaining bidders get their respective demands and the bidder who just dropped out gets nothing; (ii) if the aggregate demand is greater than the supply, then the clock increases again.

Each bidder's decision at each moment is essentially whether (i) to outbid the current prevailing price and get his/her entire demand quantity or (ii) to get only the residual quantity, which is the difference between the supply and the aggregate demand of others, at a low price. When the aggregate demand of others at the current prevailing price is not sufficient to cover the supply, it may sometimes be advantageous for a bidder to give up some of his/her demand and get the remaining demand at a low price (say, zero) rather than to get his/her entire demand at a higher price. In fact, a low price equilibrium is the only possible equilibrium when no bidder's quantity constraint is large enough to cover the supply.

Demand reduction and low price equilibrium in multi-unit auctions have attracted some attention in the literature. Noussair (1995), Engelbrecht-Wiggans and Kahn (1998) and Ausubel {\it et al.\/} (2014) have shown that a bidder has an incentive to shade his/her bid for units except for the first unit, resulting in a low equilibrium price. We note that the reason for this phenomenon in the current paper is different from the one in those papers since a bidder can submit only one lumpy bid for all units, not several different bids, in our setting.\note{On the practical side, some, if not many, markets adopt this single-bid rule. Dormady and Healy (2019) state in their footnote 2 that `It is important to note that in operating emissions markets in the U.S. firms do not submit a schedule of bids, and instead a single price-quantity bid.' Homberg and Wolak (2018) state on page 996 that `... the Colombian electricity market, where each supplier chooses one offer price for the entire capacity of each generation unit.'} Tenorio (1999) has considered lumpy bids in a setting with three identical units of a good and two symmetric bidders. In comparison, our iterative procedure systematically finds an equilibrium outcome for arbitrary units of a good and any number of asymmetric bidders. Wilson (1979), Back and Zender (1993) and Wang and Zender (2002) have found similar phenomena in share auctions under the common-value postulate instead of the private-value postulate considered in this paper and the papers cited above.

Several papers on wholesale electricity markets including von der Fehr and Harbord (1993), Fabra {\it et al.\/} (2006), Schwenen (2015) and Anderson and Holmberg (2018) study procurement auctions in which suppliers (generators) with capacity constraints compete to satisfy the buyer's electricity demand. Since the procurement auction where many sellers submit offers to a buyer is the mirror image of the auction where many buyers submit bids to a seller, our analysis can be applied equally well to this environment. The main analysis in these papers deals with either the two-bidder case or the symmetric-bidder case. In comparison, we allow many asymmetric bidders and we present an ascending auction.

The plan of the paper is as follows. The next section presents an analysis of the two-bidder case to pave the way to subsequent materials smoothly. Section 3 contains the main characterization: It introduces an iterative procedure for finding an equilibrium outcome as well as an ascending auction that has this outcome as a dominant strategy equilibrium. Section 4 concludes.
\ve

\Section{The two-bidder case}

Consider an auction in which $m$ units of a good are up for sale to two bidders. Bidder $i$ for $i=1, 2$ has a constant marginal value $v_i > 0$ up to his/her quantity constraint $q_i > 0$, but attaches no value afterward. Assume without loss of generality that $v_1 \geq v_2$. Assume also that $q_1 + q_2 > m$ since the problem is trivial otherwise.

Each bidder submits one bid for his/her demand. Let $b_i$ denote bidder $i$'s bid. The auction adopts the uniform pricing rule and the price is set to the lowest winning bid (a.k.a. the last accepted bid). Hence, when $b_1 > b_2$, the price is set to $b_1$ if $q_1 \geq m$, and set to $b_2$ if $q_1 < m$. The case when $b_1 < b_2$ can be similarly described. Lastly, when $b_1 = b_2$, the price is set to this common bid of $b_1 = b_2$.

We note that this framework is similar to that of Fabra {\it et al.} (2006) which studies procurement auctions. In particular, that paper provides a characterization of equilibria for the two-bidder case, although it is not as procedural as the approach in the present paper.

We divide the cases according to whether a bidder's demand can fill the whole supply.

\noindent [1] $q_1 \geq m$ and $q_2 \geq m$:

Since $q_2 \geq m$, which means that bidder 1 can get nothing if his/her bid is lower than bidder 2's, bidder 1 has an incentive to bid above $b_2$ whenever $b_2 < v_1$. Likewise, bidder 2 has an incentive to bid above $b_1$ whenever $b_1 < v_2$. Obviously, bidder 2 does not bid above $v_2$ because doing so results in a negative payoff. Therefore, an equilibrium bid profile is $b_1 = b_2 = v_2$. The equilibrium price is $p^*=v_2$ and the fulfilled demands are $q_1^*=m$ and $q_2^*=0$.\note{It is well-known that a pure strategy Nash equilibrium may not exist when the set of possible bids is not discrete, for instance, an interval in $\Re_+$. This can be easily overcome if we introduce the smallest money unit, say $\e > 0$. This makes the set of possible bids as a discrete set, thus avoiding the problems due to the presence of indifference. Hence, the equilibrium when $v_1 > v_2$ is that $b_2 = v_2$ and $b_1 = v_2 + \e$ because bidder 1 is ready to outbid bidder 2 and obtain all units of the good. Since we can take $\e$ arbitrarily small, the equilibrium becomes $b_1 = b_2 = v_2$ with the understanding that bidder 1 has a priority over bidder 2 in obtaining all the units he/she demands. We will follow this convention throughout the paper.}

\smallskip
\ve
\noindent [2] $q_1 \geq m$ and $q_2 < m$:

Since $q_1 \geq m$, bidder 2 has an incentive to bid above $b_1$ whenever $b_1 < v_2$, but does not bid above $v_2$. On the other hand, bidder 1 compares $m(v_1 - b_2)$, the payoff from overbidding $b_2$ and getting $m$ units, with $(m-q_2) v_1$, the payoff from bidding zero and getting $m-q_2$ units at the price of zero. Hence, given $b_2$, bidder 1 bids $b_2 + \e$ if $b_2 <  q_2 v_1/m$ and bids zero if $b_2 \geq  q_2 v_1/m$. Therefore, an equilibrium bid profile is (i) $b_1 = b_2 = v_2$ if $v_2 < q_2 v_1/m$, and (ii) $b_1 = 0$, and $b_2 = q_2 v_1/m$ or higher if $v_2 \geq q_2 v_1/m$. The equilibrium price and the fulfilled demands are $p^*=v_2, q_1^*=m$ and $q_2^*=0$ in the former, and $p^*=0, q_1^*=m-q_2$ and $q_2^*=q_2$ in the latter.

\smallskip
\noindent [3] $q_1 < m$ and $q_2 \geq m$:

Since $q_2 \geq m$, bidder 1 has an incentive to bid above $b_2$ whenever $b_2 < v_1$, but does not bid above $v_1$. On the other hand, bidder 2 compares $m(v_2 - b_1)$, the payoff from overbidding $b_1$ and getting $m$ units, with $(m-q_1) v_2$, the payoff from bidding zero and getting $m-q_1$ units at the price of zero. Hence, given $b_1$, bidder 2 bids $b_1 + \e$ if $b_1 < q_1 v_2/m$ and bids zero if $b_1 \geq q_1 v_2/m$. We note that $v_1 > q_1 v_2/m$ in this case since $v_1 \geq v_2$ and $q_1 <  m$. Therefore, an equilibrium bid profile is $b_1 = q_1 v_2/m$ or higher, and $b_2 = 0$. The equilibrium price and fulfilled demands are $p^*=0, q_1^*=q_1$ and $q_2^*=m-q_2$.

\noindent [4] $q_1 < m$ and $q_2 < m$:

Bidder 1 compares $q_1(v_1 - b_2)$, the payoff from overbidding $b_2$ and getting $q_1$ units, with $(m-q_2)v_1$, the payoff from bidding zero and getting $m-q_2$ units at the price of zero. Hence, given $b_2$, bidder 1 bids $b_2 + \e$ if $b_2 < (q_1 + q_2 - m) v_1/q_1$ and bids zero otherwise. Likewise, bidder 2 compares $q_2(v_2 - b_1)$, the payoff from overbidding $b_1$ and getting $q_2$ units, with $(m-q_1) v_2$, the payoff from bidding zero and getting $m-q_1$ units at the price of zero. Hence, given $b_1$, bidder 2 bids $b_1 + \e$ if $b_1 < (q_1 + q_2 - m)v_2/q_2$ and bids zero otherwise. Therefore, an equilibrium bid profile is (i) $b_1 = (q_1 + q_2 - m)v_2/q_2$ or higher, and $b_2 = 0$ if $(q_1 + q_2 - m)v_1/q_1 > (q_1 + q_2 - m) v_2/q_2$ or equivalently $v_1/q_1 > v_2/q_2$, and (ii) $b_1 = 0$, and $b_2 = (q_1 + q_2 -m)v_1/q_1$ or higher if $(q_1 + q_2 - m)v_1/q_1 < (q_1 + q_2 - m) v_2/q_2$ or equivalently $v_1/q_1 < v_2/q_2$. The equilibrium price and fulfilled demands are $p^*=0, q_1^*=q_1$ and $q_2^*=m-q_1$ in the former, and $p^*=0, q_1^*=m-q_2$ and $q_2^*=q_2$ in the latter.

Observe that, for the case when the bidders are symmetric such that $v_1 = v_ 2 = v$ and $q_1 = q_2 = q$, (i) if $q \geq m$, then each bidder bids $v$, and (ii) if $q < m$, then one bidder bids $(2q-m)v/q$ or higher and the other bidder bids zero.

This analysis demonstrates that a bidder has an incentive to reduce his/her demand so as to obtain the residual demand at the price of zero when the other bidder's demand does not cover the supply. It also shows that a bidder with a higher marginal value may obtain less units of the good than a bidder with a lower marginal value. For instance, when $m=1, q_1=1, q_2=0.6, v_1=1$, and $v_2=0.7$, bidder 1 gets 0.4 units whereas bidder 2 gets 0.6 units.

\Section{The main characterization}

There are $m$ units of a good and $n$ bidders in an auction. Bidder $i$ for $i = 1, \ldots, n$ has a constant marginal value $v_i > 0$ up to his/her quantity constraint $q_i > 0$, but attaches no value afterward. Assume without loss of generality that $v_1 \geq v_2 \geq \cdots \geq v_n$. Assume also that $\sum_{i=1}^n q_i > m$ since the problem is trivial otherwise. Define $\bar q_i = \min \{ q_i, m\}$, which is the maximum quantity that bidder $i$ might obtain. Let $b_i$ denote bidder $i$'s bid. Each bidder submits one bid for all his/her demand of $\bar q_i$ units.

The auction adopts the uniform pricing rule. Thus, the bids are ordered in a decreasing order to form a demand curve and the price is determined at the bid level that the demand curve intersects the supply curve, which is a vertical line at the quantity level of $m$. In particular, we set the price to the lowest winning bid (a.k.a. the last accepted bid), which corresponds to the highest price among the prices at which the induced stepwise demand curve and the vertical supply curve intersect.

We note that the price may alternatively be set to the highest losing bid (a.k.a. first rejected bid). In fact, the price is set to the highest losing bid in most papers studying the uniform price auction under incomplete information.\note{A recent exception is Burkett and Woodward (2020).} On the other hand, the price is often set to the lowest winning bid in many real-world markets including the wholesale electricity markets and the treasury securities market. The analysis below does not change in an essential way when we set the price to the highest losing bid instead of the lowest winning bid: We only have to change the inequalities regarding $q_i$ versus $m$ from strong (weak) inequalities to weak (strong) inequalities.

\smallskip
\noindent {\sl 3.1 \ \ An iterative procedure for finding an equilibrium}

Let us assume that both $v_i$ and $q_i$ for $i = 1, \ldots, n$ are common knowledge among bidders. This assumption may not be unreasonable for markets in which the same set of well-established bidders interact frequently. We also assume that the seller knows $q_i$'s throughout the paper. We can find the equilibrium by the following procedure.

\smallskip

\noindent {\bf \underbar{Step 1}:}

Given the current prevailing price of $b$, each bidder $i$ compares $\bar q_i(v_i - b)$, the payoff from overbidding $b$ and getting $\bar q_i$ units, with $(m - \sum_{j \ne i} \bar q_j) v_i$, the payoff from bidding zero and getting $m - \sum_{j \ne i} \bar q_j$ units at the price of zero, whenever $m - \sum_{j \ne i} \bar q_j > 0$. Let us define
$$\bar b_i^1  = \frac{(\sum_{j=1}^n \bar q_j - m) v_i}{\bar q_i} = v_i + \frac{(\sum_{j \ne i} \bar q_j - m) v_i}{\bar q_i}.$$
Observe that $\bar b_i^1$ is the maximum value of $b$ for which the inequality $\bar q_i(v_i - b) \geq (m - \sum_{j \ne i} \bar q_j) v_i$ holds. In addition, each bidder $i$ does not bid above $v_i$ obviously. Hence, bidder $i$'s maximum bid can be defined as
$$\hat b_i^1  = \cases{v_i & if $v_i \leq \bar b_i^1$; \cr
		    \bar b_i^1 & if $v_i > \bar b_i^1$.}$$
That is, $\hat b_i^1$ is the minimum of $v_i$ and  $\bar b_i^1$. We have:

\lemma1 $\hat b_i^1 = v_i$ if and only if $\sum_{j \ne i} q_j \geq m$. \ok

\pf See Appendix.\endpf

\noindent In other words, we have $v_i \leq \bar b_i^1$ if and only if the aggregate demand of others is not less than the supply. Let $\hat b_{(1)}^1 \geq \hat b_{(2)}^1 \geq \cdots \geq \hat b_{(n)}^1$ be the rank order of $\hat b_1^1, \ldots, \hat b_n^1$. We divide the cases.

\noindent \ [1]  When $\hat b_{(n)}^1 = \bar b_{(n)}^1$:

In equilibrium, bidder $i$ with $\hat b_i^1 = \hat b_{(n)}^1$ bids zero and others bid $\hat b_{(n)}^1$ or higher. Bidder $i$ gets $m - \sum_{j \ne i} q_j$ units and, for $j \ne i$, bidder $j$ gets $q_j$ units. The procedure stops.

\noindent \ [2] When $\hat b_{(n)}^1 = v_{(n)}$:

We claim that $\hat b_{(n)}^1 = v_n$ if $\hat b_{(n)}^1 = v_{(n)}$. To see this, observe first that we cannot have $\hat b_{(n)}^1 = v_{(n)} < v_n$ since it is a contradiction to our convention that $v_1 \geq v_2 \geq \cdots \geq v_n$. Next, suppose $\hat b_{(n)}^1 > v_n$. This implies $\hat b_n^1 > v_n$, which is again a contradiction since $\hat b_n^1 = \min \{v_n, \bar b_n^1 \} \leq v_n$. This proves the claim. Assume without loss of generality that bidder $i$ with $\hat b_i^1 = \hat b_{(n)}^1$ is bidder $n$. (Rename the bidders if necessary.) Observe that bidder $i$ must be bidder $n$ when $v_{n-1} > v_n$. Observe also that we have $\sum_{j=1}^{n-1} q_j \geq m$ by Lemma 1. We subdivide the cases.

\itemitem{[2-1]} When $\sum_{j=1}^{n-1} q_j = m$: The equilibrium is $b_1 = b_2 = \cdots = b_{n-1} = v_n$ or higher and $b_n = v_n$.  Bidder $i$ for $i = 1, \ldots, n-1$ gets $q_i$ units and bidder $n$ gets nothing. The procedure stops.

\itemitem{[2-2]} When $\sum_{j=1}^{n-1} q_j > m$: We drop bidder $n$ and move to the next step.\note{Intuitively, bidder $n$ drops out at the moment when the price exceeds $v_n$.}

\medskip
Generally, after $k$ steps with bidders $n, n-1, \ldots, n-k+1$ having dropped out, we have:

\smallskip
\noindent {\bf \underbar{Step k+1}:}

We consider the cases when $k+1 < n$ so that there are at least 2 remaining bidders: The case when $k+1 = n$ is considered below. The remaining bidders are bidders $1, \ldots, n-k$, with $v_1 \geq \cdots \geq v_{n-k}$ and $\sum_{j=1}^{n-k} q_j > m$.
\ve

Given the current prevailing price of $b$, each bidder $i$ compares $\bar q_i(v_i - b)$, the payoff from overbidding $b$ and getting $\bar q_i$ units, with $(m - \sum_{j=1, j \ne i}^{n-k} \bar q_j) (v_i - v_{n-k+1})$, the payoff from bidding (slightly above) $v_{n-k+1}$ and getting $m - \sum_{j=1, j \ne i}^{n-k} \bar q_j$ units, which is the sum of $\bar q_j$'s for $j = 1, \ldots, n-k$ except for $\bar q_i$, at the price of $v_{n-k+1}$, whenever $m - \sum_{j=1, j \ne i}^{n-k} \bar q_j > 0$. Note that the price is $v_{n-k+1}$ since bidder $n-k+1$ is ready to overbid any bid below $v_{n-k+1}$. Let us define
$$\eqalign{\bar b_i^{k+1} & = \frac{\bigl(\sum_{j=1}^{n-k} \bar q_j - m \bigr)v_i + \bigl(m - \sum_{j=1, j \ne i}^{n-k} \bar q_j \bigr) v_{n-k+1}}{\bar q_i} \cr
& = v_i + \frac{\bigl(\sum_{j=1, j \ne i}^{n-k} \bar q_j - m \bigr)\bigl(v_i -v_{n-k+1}\bigr)}{\bar q_i}.}$$
Observe that $b_i^{k+1}$ is the maximum value of $b$ for which the inequality $\bar q_i(v_i - b) \geq (m - \sum_{j=1, j \ne i}^{n-k} \bar q_j) (v_i - v_{n-k+1})$ holds. In addition, each bidder $i$ does not bid above $v_i$ obviously. Hence, bidder $i$'s maximum bid can be defined as
$$\hat b_i^{k+1} = \cases{v_i & if $v_i \leq \bar b_i^{k+1}$; \cr
		    \bar b_i^{k+1} & if $v_i > \bar b_i^{k+1}$.}$$
That is, $\hat b_i^{k+1}$ is the minimum of $v_i$ and $\bar b_i^{k+1}$. Similar to Lemma 1, we have $\hat b_i^{k+1} = v_i$ if and only if $\sum_{j=1, j \ne i}^{n-k} q_j \geq m$ when there are $n-k$ bidders, i.e., we have $v_i \leq \bar b_i^{k+1}$ if and only if the aggregate demand of others is not less than the supply. We also have:

\lemma2 $\hat b_i^{k+1} \geq v_{n-k+1}$ for all $i = 1, \ldots, n-k$. \ok

\pf See Appendix.\endpf

Let $\hat b_{(1)}^{k+1} \geq \hat b_{(2)}^{k+1} \geq \cdots \geq \hat b_{(n-k)}^{k+1}$ be the rank order of $\hat b_1^{k+1}, \ldots, \hat b_{n-k}^{k+1}$. We divide the cases.

\noindent \ [1]  When $\hat b_{(n-k)}^{k+1} = \bar b_{(n-k)}^{k+1}$:

In equilibrium, bidder $i$ with $\hat b_i^{k+1} = \hat b_{(n-k)}^{k+1}$ bids $v_{n-k+1}$ and others bid $\hat b_{(n-k)}^{k+1}$ or higher. Bidder $i$ gets $m - \sum_{j=1, j \ne i}^{n-k} q_j$ units and, for $j =1, \ldots, n-k$ and $j \ne i$, bidder $j$ gets $q_j$ units. Note that bidders $n-k+1, n-k+2, \ldots, n$ get nothing. The procedure stops.

\noindent \ [2]  When $\hat b_{(n-k)}^{k+1} = v_{(n-k)}$:

Similar to the argument in Step 1, it is straightforward to see that $\hat b_{(n-k)}^{k+1} = v_{n-k}$. Assume without loss of generality that bidder $i$ with $\hat b_i^{k+1} = \hat b_{(n-k)}^{k+1}$ is bidder $n-k$. (Rename the bidders if necessary.) Observe that bidder $i$ must be bidder $n-k$ when $v_{n-k-1} > v_{n-k}$. Observe also that we have $\sum_{j=1}^{n-k-1} q_j \geq m$ by a reasoning similar to Lemma 1. We subdivide the cases.

\itemitem{[2-1]} When $\sum_{j=1}^{n-k-1} q_j = m$: The equilibrium is $b_1 = b_2 = \cdots = b_{n-k-1} = v_{n-k}$ or higher and $b_{n-k} = v_{n-k}$. Bidder $i$ for $i = 1, 2, \ldots, n-k-1$ gets $q_i$ units and bidder $i$ for $i = n-k, n-k+1, \ldots, n$ gets nothing. The procedure stops.

\itemitem{[2-2]} When $\sum_{j=1}^{n-k-1} q_j > m$: We drop bidder $n-k$ and move to step $k+2$.\note{Intuitively, bidder $n-k$ drops out at the moment when the price exceeds $v_{n-k}$.}

Note that, by defining $v_{n+1} = 0$ as a convention, step $k+1$ applies to all $k = 0, 1, \ldots, n-2$. Lastly,

\smallskip
\noindent {\bf \underbar{Step n}:}

The only remaining bidder is bidder 1. In equilibrium, bidder 1 gets $m$ units of the good at a price $v_2$.\note{Recall our convention stated in the previous section to avoid the problems due to the presence of indifference.}
\medskip

Observe that the equilibrium strategies are described in case [1] and case [2-1]. We summarize the resulting equilibrium price and fulfilled demands as follows.

\prop1 Suppose the procedure stops at step $k+1$ for $k = 0, \ldots, n-2$.\note{Recall that the bidders at the beginning of this step are bidders $1, \ldots, n-k$.}\it The equilibrium price is given as
$$p^* = \cases{v_{n-k+1} & if \ \ $\sum_{j=1, j \ne i}^{n-k} q_j < m$; \cr
               v_{n-k} & if \ \ $\sum_{j=1, j \ne i}^{n-k} q_j = m$.}$$
The fulfilled demands are $q_i^* = m - \sum_{j=1, j \ne i}^{n-k} q_j$ for bidder $i$ with $\hat b_i^{k+1} = \hat b_{(n-k)}^{k+1}$; $q_j^* = q_j$ for bidder $j = 1, \ldots, n-k$ and $j \ne i$; and $q_j^*=0$ for bidder $j = n-k+1, n-k+2, \ldots, n$. If the procedure proceeds up to step $n$, the equilibrium price is given as $p^* = v_2$ and the fulfilled demands are $q_1^*=m$ and $q_j^*=0$ for $j = 2, \ldots, n$. \ok

This proposition shows that the equilibrium price is always set at one of the bidders' marginal values (or at zero). Observe however that this does not imply that bidders are allocated the good according to the decreasing order of marginal values. That is, a bidder with a higher marginal value may obtain less units of the good than a bidder with a lower marginal value: The marginal value of bidder $i$ with $\hat b_i^{k+1} = \hat b_{(n-k)}^{k+1}$ may be higher than the marginal value of bidder $j$ for $j = 1, \ldots, n-k$, but bidder $i$ may obtain less units than bidder $j$.\note{A simple example is given at the end of the last section.} The reason is that, if $\sum_{j=1, j \ne i}^{n-k} q_j < m$ holds at step $k+1$ for $k = 0, \ldots, n-2$, then case [1] is in effect and a low price equilibrium where bidder $i$ gives up his/her full demand but only obtains the residual supply with a bid of $v_{n-k+1}$ occurs. Observe that the bid $v_{n-k+1}$ is lower than any remaining bidder's marginal value $v_j$ for $j = 1, \ldots, n-k$. In essence, inefficiency results from demand reduction and low price equilibrium. As a matter of fact, this equilibrium is inevitable if no bidder's quantity constraint is large enough to cover the whole supply.

\prop2 If $q_i < m$ for all $i = 1, \ldots, n$, then a low price equilibrium is the only possible equilibrium. \ok

\pf Suppose not. This implies that $\sum_{j=1, j \ne i}^{n-k} q_j \geq m$ holds for all $i = 1, \ldots, n-k$ and for all step $k+1$ for $k = 0, \ldots, n-2$. In particular, we have $q_1 \geq m$ and $q_2 \geq m$ in step $n-1$ with 2 bidders remaining. This is a contradiction to the fact that $q_i < m$ for all $i = 1, \ldots, n$. \endpf

The procedure can be summarized by the following algorithm.

\bigskip
\hrule
\medskip

\noindent {\bf Algorithm} Equilibria of Uniform Price Auctions with Quantity Constraints

\smallskip
\hrule
\medskip

\noindent {\bf Input:} $m, \{v_i, q_i\}_{i=1}^n$, $v_{n+1}=0$

\noindent {\bf Output:} $p^*, \{b_i^*, q_i^*\}_{i=1}^n$

\noindent \ \ \phantom{0}1: \ $\bar q_i = \min\{q_i,m\}$

\noindent \ \ \phantom{0}2: \ $k=0$

\noindent \ \ \phantom{0}3: \ {\bf while} $k < n-1$ {\bf do}

\noindent \ \ \phantom{0}4: \ \ \ \ \ {\bf for} $i = 1, \ldots, n-k$ {\bf do}

\noindent \ \ \phantom{0}5: \ \ \ \ \ \ \ \ \ $\bar b_i^{k+1} = v_i + \bigl(\sum_{j=1, j \ne i}^{n-k} \bar q_j - m \bigr)\bigl(v_i -v_{n-k+1}\bigr)/\bar q_i$

\noindent \ \ \phantom{0}6: \ \ \ \ \ \ \ \ \ $\hat b_i^{k+1} = v_i$ {\bf if} $v_i \leq \bar b_i^{k+1}$ {\bf else} $\bar b_i^{k+1}$

\noindent \ \ \phantom{0}7: \ \ \ \ \ {\bf end for}

\noindent \ \ \phantom{0}8: \ \ \ \ \ Arrange $\hat b_i^{k+1}$'s in a decreasing order, yielding $\hat b_{(1)}^{k+1} \geq \hat b_{(2)}^{k+1} \geq \cdots \geq \hat b_{(n-k)}^{k+1}$

\noindent \ \ \phantom{0}9: \ \ \ \ \ {\bf if} $\hat b_{(n-k)}^{k+1} = \bar b_{(n-k)}^{k+1}$

\noindent \ \ 10: \ \ \ \ \ \ \ \ \ return $p^*, \{b_i^*, q_i^*\}_{i=1}^n$ as given in case [1]

\noindent \ \ 11: \ \ \ \ \ \ \ \ \ {\bf break}

\noindent \ \ 12: \ \ \ \ \ {\bf else} \ \ \ \ \ \ \ \ \ \ \ \ \ \ \ \ \ \ \ \ \ $\#$ that is, $\hat b_{(n-k)}^{k+1} = v_{(n-k)}$

\noindent \ \ 13: \ \ \ \ \ \ \ \ \ {\bf if} $\sum_{j=1}^{n-k-1} q_j = m$

\noindent \ \ 14: \ \ \ \ \ \ \ \ \ \ \ \ \ return $p^*, \{b_i^*, q_i^*\}_{i=1}^n$ as given in case [2-1]

\noindent \ \ 15: \ \ \ \ \ \ \ \ \ \ \ \ \ {\bf break}

\noindent \ \ 16: \ \ \ \ \ \ \ \ \ {\bf else} \ \ \ \ \ \ \ \ \ \ \ \ \ \ \ \ \ $\#$ that is, $\sum_{j=1}^{n-k-1} q_j > m$

\noindent \ \ 17: \ \ \ \ \ \ \ \ \ \ \ \ \ drop bidder $n-k$

\noindent \ \ 18: \ \ \ \ \ \ \ \ \ \ \ \ \ $k \leftarrow k+1$

\noindent \ \ 19: \ \ \ \ \ \ \ \ \ {\bf end if}

\noindent \ \ 20: \ \ \ \ \ {\bf end if}

\noindent \ \ 21: \ {\bf end while}

\noindent \ \ 22: \ {\bf if} $k=n-1$

\noindent \ \ 23: \ \ \ \ \ return $p^*, \{b_i^*, q_i^*\}_{i=1}^n$ as given in step n

\smallskip
\hrule

\bigskip

We provide examples that illustrate how the procedure works.

\eg1 Suppose there are three units of a good, i.e., $m=3$, and three bidders with $(v_1, v_2, v_3) = (0.7, 0.5, 0.3)$ and $(q_1, q_2, q_3)= (3, 2, 3)$. In step 1, we have $\bar b_1 = 1.17, \hat b_1 = 0.7, \bar b_2 = 1.25, \hat b_2 = 0.5, \bar b_3 = 0.5$ and $\hat b_3 = 0.3$. We drop bidder 3. In step 2, we have $\bar b_1 = \hat b_1 = 0.57, \bar b_2 = 0.5$ and $\hat b_2 = v_2 = 0.5$. We drop bidder 2. In equilibrium, bidders 1 and 2 bid 0.5 and bidder 3 bids 0.3. The equilibrium price is 0.5. Bidder 1 gets 3 units, and bidders 2 and 3 get nothing.

Observe that bidder 1 gets 3 units while bidder 2 gets nothing, even though both bid 0.5. This holds since the procedure drops bidder 2 in the previous step. As noted in footnote 4, we may think of this as bidder 1 bidding slightly above 0.5 to win over bidder 2. A similar observation applies to the next example as well.

\eg2 Suppose there are three units of a good, i.e., $m=3$, and three bidders with $(v_1, v_2, v_3) = (1.0, 0.5, 0.1)$ and $(q_1, q_2, q_3)= (2, 2, 1)$. In step 1, we have $\bar b_1 = 1.0, \hat b_1 = 1.0, \bar b_2 = 0.5, \hat b_2 = 0.5$, $\bar b_3 = 0.2$ and $\hat b_3 = 0.1$. We drop bidder 3. In step 2, we have $\bar b_1 = \hat b_1 = 0.55$ and $\bar b_2 = \hat b_2 = 0.3$. In equilibrium, bidder 1 bids 0.3 or higher, and bidders 2 and 3 bid 0.1. The equilibrium price is 0.1. Bidder 1 gets 2 units, bidder 2 gets 1 unit, and bidder 3 gets nothing. \ok

There exists another equilibrium in the previous example: Bidder 1 bids 0.1, bidder 2 bids 0.55, and bidder 3 bids 0.1. The equilibrium price is 0.1. Bidder 1 gets 1 unit, bidder 2 gets 2 units, and bidder 3 gets nothing. Note that the good is not allocated efficiently in this equilibrium. That is, bidder 1 who has the highest value gets only one unit of the good.

Hence, the equilibrium outcome, that is, the equilibrium allocation of items to bidders and the equilibrium price, found by our procedure may not be unique. However, we show that this is the only dominant strategy equilibrium outcome of an ascending auction defined below.

\smallskip
\noindent {\sl 3.2 \ \ A dynamic implementation}
\smallskip

Assume now that bidders have private information regarding their respective values, i.e., each $v_i$ is known only to bidder $i$. We continue to assume, on the other hand, that the quantity constraints $q_i$'s are common knowledge among all parties including the seller. We introduce the English auction with quantity constraints, which is an open ascending auction and which incorporates the quantity constraints to the standard English auction.

There is a clock showing the current price, which continuously increases over time. The clock starts at zero, and all bidders participate initially. As the price increases, a bidder may drop out. Let $N(0)$ be the initial set of bidders and $N(b)$ be the set of remaining (i.e., active) bidders at a price $b$.\note{$N(b)$ as a function of the current price $b$ is a right-continuous function.} In addition, the auction keeps track of the provisional price $\hat p$, which is set to zero initially.

When a bidder drops out, the clock stops temporarily and the auctioneer checks whether the aggregate demand of the remaining bidders is greater than or equal to the supply. Suppose bidder $i$ drops out at a price $b$.\note{When there is more than one bidder dropping out simultaneously, we can choose one bidder in any fashion and apply the procedure in the text while other bidders are entitled to remain active. Note in particular that if the auction continues just after dropping the bidder (that is, subcase [2-2] below holds), then the procedure in the text is applied promptly again for the remaining bidders.} The auctioneer checks whether $\sum_{j \in N(b)} \bar q_j \geq m$.
\item{[1]} If $\sum_{j \in N(b)} \bar q_j < m$ holds, then the auction ends: The final price is set to $\hat p$, bidder $i$ gets $m - \sum_{j \in N(b)} q_j$ units and, for $j \in N(b)$, bidder $j$ gets $q_j$ units.
\item{[2]} If $\sum_{j \in N(b)} \bar q_j \geq m$ holds, then the provisional price $\hat p$ is updated to $b$.
\itemitem{[2-1]} If $\sum_{j \in N(b)} \bar q_j = m$ holds, the auction ends: The final price is set to $\hat p$, which has the updated value of $b$, bidder $i$ gets nothing and, for $j \in N(b)$, bidder $j$ gets $q_j$ units.
\itemitem{[2-2]} If $\sum_{j \in N(b)} \bar q_j > m$ holds, then the clock increases again.

A bidder's strategy is a mapping from his/her value and the bidding history to the decision to drop out. Formally, bidder $i$'s (pure) strategy at time $t$ when he/she is still active is $s_i(v_i, h_t)$, where $v_i$ is bidder $i$'s value and $h_t$ is a bidding history at the beginning of time $t$. The action $s_i(v_i, h_t)$ belongs to the set $\{0, 1\}$, where $0$ stands for `drop out' and $1$ stands for `remain active'. The action is irreversible. Thus, $s_i(v_i, h_0) = 1$, and once it has dropped to $0$, it stays at $0$ forever.

The equilibrium strategy in this auction is straightforward. Let $N_t$ be the set of active bidders at the beginning of time $t$. At a history with $(\hat p, N_t, b)$, bidder $i$ with value $v_i$ compares $\bar q_i(v_i - b)$ with $(m - \sum_{j \in N_t \setminus \{i\}} \bar q_j)(v_i - \hat p)$ whenever $m - \sum_{j \in N_t \setminus \{i\}} \bar q_j > 0$. Hence, it is weakly dominant to remain active if and only if $b \leq v_i$ and
$$b \leq  v_i + \frac{\bigl( \sum_{j \in N_t \setminus \{i\}} \bar q_j - m \bigr)(v_i - \hat p)}{\bar q_i} \equiv \bar b_i(v_i, \hat p, N_t).$$

Observe the similarity of $\bar b_i(v_i, \hat p, N_t)$ to $\bar b_i^{k+1}$ of the previous subsection. We have $v_i \leq \bar b_i(v_i, \hat p, N_t)$ if and only if $\sum_{j \in N_t \setminus \{i\}} \bar q_j \geq m$. Hence, if bidder $i$ drops out at $b$ and $v_i > \bar b_i(v_i, \hat p, N_t)$ holds, then the auction ends. The equilibrium price is the provisional price $\hat p$. Bidder $i$ gets $m - \sum_{j \in N_t \setminus \{i\}} q_j$ units and, for $j \in N_t \setminus \{i\}$, bidder $j$ gets $q_j$ units. Next, if bidder $i$ drops out at $b$ and $v_i \leq \bar b_i(v_i, \hat p, N_t)$ holds, then the provisional price is updated to $b$, which is equal to $v_i$ given the equilibrium strategy: In case when $\sum_{j \in N_t \setminus \{i\}} \bar q_j = m$, the auction ends. The equilibrium price is set to $\hat p$, which has the updated value of $b$. Bidder $i$ gets nothing and, for $j \in N_t \setminus \{i\} = N(b)$, bidder $j$ gets $q_j$ units. Otherwise, the clock increases again.

Thus, the dominant strategy equilibrium of this auction is outcome equivalent to the equilibrium found by the procedure in the previous subsection. Observe that the equilibrium of the ascending auction is dominant under the assumption that bidders have private information regarding their respective values whereas the equilibrium found by previous procedure is a Nash equilibrium under complete information.

\smallskip
\noindent {\sl 3.3 \ \ Extensions}
\smallskip

It is straightforward to incorporate a positive reserve price in the analysis. It can be shown that the reserve price may increase the seller's revenue when there is insufficient competition in the sense that the aggregate demand of others is less than the supply, i.e., $\sum_{j \ne i} \bar q_j < m$. Next, while the main analysis focuses on auction environments in which many buyers compete to buy multiple units of a good, it can be applied equally well to the procurement auction environments in which many producers compete to sell multiple units of a good since the latter is the mirror image of the former. The changes we need are (i) to introduce the maximum allowable price or price cap, and (ii) to apply the previous analysis symmetrically.

If the procurement auction with capacity constraints adopts the discriminatory pricing rule instead of the uniform pricing rule, i.e., if each seller is paid his/her own offer, then the competition among bidders is equivalent to price competition in a capacity-constrained oligopoly. A well-known property of the latter is the existence of Edgeworth cycle. Thus, price may fluctuate since firms undercut each other until this price war becomes too costly, after which firms increase prices and then again the price war ensues. Levitan and Shubik (1972), Osborne and Pitchik (1986) and Davidson and Denekere (1986) have characterized equilibria for the case of two firms with the same marginal costs.

Edgeworth cycle is also observed in auctions with the discriminatory pricing rule.\note{These auctions are alternatively called as discriminatory auctions or pay-as-bid auctions.} Suppose there are two units of a good and two bidders. Let $v_1=7, v_2=5$ and $q_1=2, q_2=1$. As long as bids are below 3.5, bids escalate. When a bid reaches 3.5, bidder 1 bids zero since he/she is indifferent between getting two units at the price of 3.5 and getting one unit at the price of zero. Bidder 2 then follows suit to get one unit at a lower price. This process repeats.

\Section{Conclusion}

We have presented an iterative procedure that systematically finds a Nash equilibrium outcome of the uniform price auctions when there are many asymmetric quantity-constrained bidders. We have also presented an ascending auction that has this outcome as a dominant strategy equilibrium under the assumption that bidders have private information regarding their respective values. We have shown that demand reduction and low price equilibrium might occur in these multi-unit auctions: The reason for this phenomenon is somewhat different from the one in the previous literature since each bidder in this paper can submit only one bid for his/her entire demand so that differential bid shading of multiple bids is not possible. In fact, a low price equilibrium is the only possible equilibrium when no bidder's quantity constraint is large enough to cover the supply.

We have assumed that the bidders have constant marginal values up to his/her quantity constraints. It is a challenging future research agenda to construct a procedure that explicitly finds the equilibrium for the case when bidders have general marginal values and submit multiple bids. As we have discussed in the last subsection, the procurement auction with capacity constraints when it adopts the discriminatory pricing rule is strategically equivalent to price competition in a capacity-constrained oligopoly. It is also a challenging future research agenda to examine this kind of competition for the case when the sellers have asymmetric marginal costs and/or when there are more than two sellers.

\bigskip
\noindent {\bf Acknowledgments:} I thank anonymous reviewers for many helpful comments and suggestions.

\appx

\pfo{Lemma 1} Assume that $\hat b_i^1 = v_i$. This is equivalent to $\sum_{j \ne i} \bar q_j \geq m$. Since $\bar q_j \leq q_j$ for all $j$, we get $\sum_{j \ne i} q_j \geq m$.

Assume next that $\hat b_i^1 = \bar b_i^1$. This is equivalent to $\sum_{j \ne i} \bar q_j < m$. We have $\bar q_j = q_j$ for all $j \ne i$ since otherwise, i.e., if $\bar q_j = m$ for some $j \ne i$, then we must have $\sum_{j \ne i} \bar q_j \geq m$. Thus, we get $\sum_{j \ne i} q_j < m$. \endpf

\pfo{Lemma 2} Obvious if $\hat b_i^{k+1} = v_i$ since $v_i \geq v_{n-k+1}$ for all $i = 1, \ldots, n-k$. So, assume that $\hat b_i^{k+1} = \bar b_i^{k+1}$. We have $\bar b_i^{k+1} - v_{n-k+1} = (\sum_{j=1}^{n-k} \bar q_j - m)(v_i - v_{n-k+1})/\bar q_i \geq 0$ since $\sum_{j=1}^{n-k} \bar q_j > m$. To see that the last inequality holds, suppose for the sake of contradiction that $\sum_{j=1}^{n-k} \bar q_j \leq m$. We cannot have $\bar q_j = m$ for any $j = 1, \ldots, n-k$ since there are at least 2 remaining bidders. Hence, $\bar q_j = q_j$ for all $j = 1, \ldots, n-k$ and we have $\sum_{j=1}^{n-k} q_j \leq m$, which contradicts the fact that $\sum_{j=1}^{n-k} q_j > m$. \endpf

\ref

\wp{ACER}{2022}{ACER's final assessment of the EU wholesale electricity market design}{European Union}

\paper{Anderson, E., Holmberg, P.}{2018}{Price stability in multi-unit auctions}{\jet 175}{318-341}

\paper{Ausubel, L., Cramton, P., Pycia, M., Rostek, M., Weretka, M.}{2014}{Demand reduction and inefficiency in multi-unit auctions}{\res 81}{1366-1400}

\paper{Back, K., Zender, J.}{1993}{Auctions of divisible goods: On the rationale for the treasury experiment}{{\it Review of Financial Studies\/} 6}{733-764}

\paper{Burkett, J., Woodward, K.}{2020}{Uniform price auctions with a last accepted bid pricing rule}{\jet 185}{Article 104954}

\paper{Cramton, P., Kwerel, E., Rosston, G., Skrzypacz, A.}{2011}{Using spectrum auctions to enhance competition in wireless services}{\jle 54}{S167-S188}

\paper{Davidson, C., Deneckere, R.}{1986}{Long-run competition in capacity, short-run competition in price, and the Cournot model}{\rje 17}{404-415}

\paper{Dormady, N., Healy, P.}{2019}{The consignment mechanism in carbon markets: A laboratory investigation}{{\it Journal of Commodity Markets\/} 14}{51-65}

\paper{Engelbrecht-Wiggans, R., Khan, C.}{1998}{Multi-unit auctions with uniform prices}{\et 12}{227-258}

\paper{Fabra, N., von der Fehr, N-H.M., Harbord, D.}{2006}{Designing electricity auctions}{\rje 37}{23-46}

\paper{Homberg, P., Wolak, F.}{2018}{Comparing auction designs where suppliers have uncertain costs and uncertain pivotal status}{\rje 49}{995-1027}

\paper{Horta\c{c}su, A., Kastl, J.,  Zhang, A.}{2018}{Bid shading and bidder surplus in the US treasury auction system}{\aer 108}{147-169}

\paper{Levitan, R., Shubik, M.}{1972}{Price duopoly and capacity constraints}{{\it International Economic Review\/} 13}{111-122}

\wp{Malvey, P., Archibald, C.}{1998}{Uniform price auctions: Update of the treasury
experience}{US Treasury}

\book{Milgrom, P.}{2004}{Putting Auction Theory to Work}{Cambridge University Press}

\paper{Noussair, C.}{1995}{Equilibria in a multi-object uniform price sealed bid auction with multi-unit demands}{\et 5}{337-351}

\paper{Osborne, M., Pitchik, C.}{1986}{Price competition in a capacity-constrained duopoly}{\jet 38}{238-260}

\paper{Schwenen, S.}{2015}{Strategic bidding in multi-unit auctions with capacity constrained bidders: the new York capacity market}{\rje 46}{730-750}

\paper{Tenorio, R.}{1999}{Multiple unit auctions with strategic price-quantity decisions}{\et 13}{247-260}

\paper{von der Fehr, N-H.M., Harbord, D.}{1993}{Spot market competition in the UK electricity industry}{\ej 103}{531-546}

\paper{Wang, J., Zender, J.}{2002}{Auctioning divisible goods}{\et 19}{673-705}

\paper{Wilson, R.}{1979}{Auctions of shares}{\qje 93}{675-689}

\bye